\documentclass{bmcart}

\usepackage{amsthm,amsmath}
\usepackage{amsmath}
\usepackage{algorithm}
\usepackage[noend]{algpseudocode}
\usepackage[utf8]{inputenc} 

\def\includegraphics{}

\startlocaldefs
\endlocaldefs

\usepackage{graphicx}
\graphicspath{ {./images/} }

\begin{document}

\begin{frontmatter}

\begin{fmbox}
\dochead{Research}


\title{A Kubernetes `Bridge' operator between
cloud and external resources.
}

\author[
  addressref={aff1},
  email={blublinsky@ibm.com}
]{\inits{B.}\fnm{Boris} \snm{Lublinsky}}
\author[
  addressref={aff2},
  corref={aff2}, 
  email={elise.jennings@ibm.com}
]{\inits{E.}\fnm{Elise} \snm{Jennings}}
\author[
  addressref={aff2},                   
  email={viktoria.spisakova@ibm.com}   
]{\inits{V.}\fnm{Viktória } \snm{Spišaková}}

\address[id=aff1]{%
  \orgdiv{IBM Systems},
  \city{Chicago, IL},
  \cny{USA}
}
\address[id=aff2]{
  \orgdiv{IBM Research - Europe},             
  \city{Dublin},                              
  \cny{Ireland}                                    
}



\end{fmbox}


\begin{abstractbox}

\begin{abstract} 
Many scientific workflows require dedicated compute resources, including HPC clusters with optimized software, quantum resources, and dedicated hardware cluster systems like Ray, for example. At the same time,  many scientific workflows today are built on Kubernetes leveraging growing support for workflow and support tools. To address the growing demand to support workflows on both cloud and dedicated compute resources we present the Bridge Operator, a software extension for container orchestration in Kubernetes which facilitates the submission and monitoring of long running processes on external systems which have their own cluster resources manager (SLURM, LSF, quantum services and Ray). The Bridge Operator consists of a custom Kubernetes controller that employs a Kubernetes Custom Resource Definition to manage applications. 
We present controller logic to manage the cloud container orchestration and external resource workload manager interface, a resource definition to submit HTTP/HTTPS requests to the external resource, and a controller pod communicating with the external resource manager to submit and manage job execution. The implementation allows us to mirror the external  resource in Kubernetes pods,  which allows the operator  to use these pods as proxies to control the external system. The implementation is agnostic to the choice of resource manager but assumes the system exposes a HTTP/HTTPS API for its control/management. The Bridge Operator automates the role of a human operator running jobs on a black box external resource as part of a complex hybrid workflow on the Cloud.

\end{abstract}




\end{abstractbox}
%

\end{frontmatter}



\section{Introduction}

Cloud-based scientific workflow platforms provide dynamically managed and monitored environments with access to a wide variety of on-demand computing resources. One focus in Cloud computing is flexibility where multiple environments can be deployed on the same node at the same time and efficient resource sharing on the nodes can in turn lead to increased resource utilization. Docker based containerization in cloud environments ensures portability and allows users to deploy applications easily among clusters. Orchestration provided by Kubernetes leads to efficient auto-scaling and provisioning of cluster resources.
On the other hand, many scientific workflows require specialized hardware (GPU’s, quantum, etc) and software to carry out complex scientific computations. Despite many attempts to extend the reach of Kubernetes to support specialized hardware and add additional resource managers it does not seem feasible, at the moment, to bring all of these resources into a single Kubernetes cluster. There are multiple issues to be considered:
\begin{itemize}
    \item There is a continuous evolution in specialized hardware and backporting Kubernetes support to them is not always easy.
\item The Kubernetes resource manager, while extremely well suited for common processing, is not optimized for HPC workloads when compared to specialized resource managers, like LSF or Slurm. 
\end{itemize}
 Dedicated computing systems like HPC clusters, are pre-configured by systems administrators with specialized software libraries and frameworks. As a result workloads for these dedicated resources must be built to be hardware specific for performance. On these systems provisioning of resources is typically carried out by a workload manager which consists of a resource manager and a job scheduler. The workload manager allocates resources, schedules the jobs, determines priorities, submits the jobs to the compute nodes and manages allocations and dependencies.
The rise of scientific machine learning and high performance data analytics has led to an increase in complex and diverse workloads which may mix Cloud-based and on-premise external resources such as a HPC cluster or  an AI accelerator. Many workflows in  scientific domains like Climate, Pharma, Material science, Astronomy, Bioinformatics, Physics, and Earth Science which would have traditionally run only on a HPC cluster, now span from edge devices and cloud-based clusters to HPC systems. There is a growing demand to accelerate workflow components by deploying jobs to advanced systems like quantum. Additionally new software implementations, for example Ray, provide additional speed up of Python-based execution and support for a wider range of hardware options.
In this paper we present a Kubernetes operator, the ‘Bridge Operator’, which allows users to dispatch jobs from a Kubernetes workflow to external resources (seen as a “black box”) and monitor the job until completion, re-submit upon failure, delete a running job, upload input data and additional files and retrieve outputs.

This paper is organized as follows: In Section \ref{sect:background} we present the background on External Workload managers in \ref{subsect:wlm}, Kubernetes cloud orchestration in \ref{subsect:ks} and Kubernetes Operators \ref{subsect:operators}. In Section \ref{sect:relatedwork} we present related work in this area and in Section \ref{sect:obj} we outline the objectives of this work. In Section \ref{sect:archntools} we describe the Bridge architecture and tools and in Section \ref{sect:workflow} we present a workflow integration. In conclusion \ref{sect:conc} we will outline current results and potential future work.

\section{Background}\label{sect:background}
In this section we describe commonly used workload managers, Slurm, LSF, Quantum, and Ray, in Section \ref{subsect:wlm}. In Section \ref{subsect:ks} we outline the cloud orchestration framework, Kubernetes, and in Section \ref{subsect:operators} we describe Kubernetes Operators.

\subsection{External workload managers}\label{subsect:wlm}
In general the compute nodes on HPC clusters can only be accessed using resource and workload manager programs which control the allocations of processors and memory as well as scheduling the jobs to be run. Job and queue scheduling can follow priority policies specific to the HPC cluster or specific to a queue in the cluster such as the size, type and duration of the job. Two workload management programs which exemplify current usage are the Slurm and IBM Spectrum LSF software packages.

Slurm \cite{slurm} is a cluster management and job scheduling program for Linux clusters. Slurm manages the allocations giving users access to the compute nodes for their jobs and controls the execution and monitoring of those jobs. Slurm also manages job queues, topology aware job allocations, reservations and backfill policies.  A slurm cluster is composed of a centralized daemon, slurmctld, running on a management node together with slurmd daemons on each of the compute nodes which executes the work. The Slurm daemons manage the compute nodes, partitions of nodes, memory allocations and jobs. An optional slurmrestd daemon is available to interact with Slurm through its REST API which is an HTTP server.
User command line tools such as sinfo provide information on the current state of the system (nodes, partitions, queues), squeue provides information on the jobs on the system and srun to create a resource allocation and run jobs.

The IBM Spectrum LSF (`LSF', short for load sharing facility) software \cite{lsf} is a workload manager which allocates resources and provides a job management framework to run and monitor HPC jobs. The core of LSF includes various daemon processes, depending on their role in the cluster. There are daemons for job requests and dispatch, scheduling and execution.
User command line tools such as bsub are used to submit jobs to a queue and specify job submission options to modify the default job behavior. Submitted jobs wait in queues until they are scheduled and dispatched to a host for execution. At job dispatch, LSF checks to see which hosts are eligible to run the job and at runtime the environment is copied from the submission host to the execution host. LSF can be accessed by users and administrators by a command-line interface, an API, or through the IBM Spectrum LSF Application Center.

IBM recently released Quantum service \cite{quantum} which run on the  IBM cloud. Although technically this is not a resource manager, these APIs provide functionality similar to APIs provided by HPC workload managers described above and can be used leveraging the same programming paradigm. Via these APIs it is possible to create jobs, monitor their execution and upload execution results to object storage.

Another interesting implementation is Ray \cite{ray} - an open source project that makes it simple to scale any compute-intensive Python workload. Ray implements a flexible distributed execution framework providing support for many types of advanced hardware. A new job submission SDK, provided by Ray allows users to  submit and monitor jobs on a Ray cluster. 

\subsection{Kubernetes}\label{subsect:ks}
Kubernetes \cite{k8s}  is a portable, extensible, open-source platform for managing containerized workloads and services at scale. In a nutshell Kubernetes is about managing resources - endpoints in the Kubernetes API that store a collection of API objects of a certain kind. The following is the list of the base kubernetes resources:
\begin{itemize}
    \item Pod. A group of one or more containers.
\item Service. An abstraction that defines a logical set of pods as well as the policy for accessing them.
\item Volume. An abstraction that allows data to persist. (This is necessary because containers are ephemeral - meaning data is deleted when the containers themselves cease to exist.)
\item Secrets and configuration maps - resources used to store confidential and non-confidential data as key-value pairs.
\item Namespace. A segment of the cluster dedicated to a certain purpose, for example a certain project or team of devs.
\item ReplicaSet (RS). Ensures that the desired number of pods are running.
\item Deployment. Offers declarative updates for pods and RS.
\item StatefulSet. A workload API object that manages stateful applications, such as databases.
\item DaemonSet. Ensures that all or some worker nodes run a copy of a pod. This is useful for daemon applications.
\item Job. Creates one or more pods, runs certain tasks to completion, then deletes the pods.
\item Etc.
\end{itemize}
Kubernetes manages containers which package code, runtime, system tools, system libraries, and configurations altogether. Containerization allows reproducible runs of an application irrespective of where it runs.
One of the strongest features of Kubernetes is that it is a declarative system - you supply the representation of the desired state to the system and the system reads the current state and determines the sequence of commands to transition to this desired state. In a Kubernetes implementation, every resource type has its own controller - the component that determines the necessary sequence of commands required for state transition. The are several advantages in such declarative approach:
\begin{itemize}
    \item It makes it significantly simpler for the deployer - instead of the user having to determine the sequence of commands to transition the system to the desired state, they only have to define the desired state.
    \item An additional advantage of the declarative system is the ability to react to any unintended state changes. In the cases of unintended state change, the system will determine and apply the set of correcting actions bringing it back to the desired state.
\end{itemize}
Pods  can be killed, replaced, and be automatically re-started. Upon re-starting the pod, a container is given access to the support volumes, secrets and configuration maps that are needed for it to function.
Another important Kubernetes feature is its extensibility through the custom resources - extensions of the Kubernetes API. Many core Kubernetes functions are now built using custom resources, making Kubernetes more modular.
Custom resources (CRs) can appear and disappear in a running cluster through dynamic registration of custom resource definitions (CRD), and cluster administrators can update custom resources independently of the cluster itself. Once a custom resource is installed, users can create and access its objects using kubectl, with a similar ease as for built-in resources.

\subsection{Kubernetes Operators}\label{subsect:operators}
In Kubernetes, controllers of the control plane implement control loops that repeatedly compare the desired state of the cluster to its actual state and reconcile any discrepancies. A Kubernetes operator is a software extension to allow for the creation of a custom controller that uses CRs to manage applications and their components. The CRs used by the Operator to manage the application contain specific high-level configuration and settings that allow easy user customization. The Operator also implements low level actions, based on the logic defined within the custom controller.
A Kubernetes operator plays the role of automating all aspects of human operators who look after specific applications and services. These Human operators have deep knowledge of how the system ought to behave (credentials, endpoints, secrets), how the application is deployed (scheduler commands), and how to react if there are problems (job termination).The original definition of operators in Kubernetes had a single application focus which included domain or application-specific knowledge to automate the entire lifecycle of the software it manages. A good example is TfJob \cite{} which is a Kubernetes operator with a custom controller and CRD to run TensorFlow training jobs on Kubernetes.

\section{Related Work} \label{sect:relatedwork}
The Kubernetes operator is well known prior art and related work has been done on the general problem of supporting compute intensive or HPC workloads on cloud resources. 
Zhou et al. \cite{DBLP:journals/corr/abs-2012-08866} proposed a hybrid architecture solution to integrate the Torque \cite{torque} workload manager with Kubernetes. This work introduces a dual level of scheduling where container orchestration on the HPC cluster can be performed by the Kubernetes orchestrator in the cloud. The proposed architecture assumes a shared login node which  belongs to both the Kubernetes and TORQUE clusters. This is a virtual node that represents a worker node in the Kubernetes cluster and acts as a login node for TORQUE. The login node only submits the TORQUE job to the HPC cluster and is not included in the TORQUE compute node list. Job submission on the HPC cluster side will not be scheduled to this login node. The Kubernetes cluster incorporates a master node that schedules the jobs to the worker nodes.

The WLM Operator \cite{wlm} is another Kubernetes operator implementation which connects a Kubernetes node with a Slurm cluster allowing Kubernetes to integrate with Slurm. Each Slurm partition (queue) is considered as a dedicated virtual node in Kubernetes and compute resources on the Slurm cluster are propagated and visible to Kubernetes. This approach implements a GRPc server in order to submit requests to the Slurm cluster. 

Both the WLM and Torque operator implementations operate using a grey box approach, where the HPC job scheduler is effectively visible to the Kubernetes applications. In both, a single instance of the operator is limited to a specific workload manager on a given external system and in this sense neither can be considered a generic approach i.e. a workflow with steps to be run on a Torque and Slurm cluster would require two different operators as well as a shared Kubernetes - Torque login node and a Kubernetes - Slurm login node to run.

Piras et al. \cite{10.1007/978-3-030-34356-9_3} propose a hybrid Kubernetes on HPC approach where the size of the Kubernetes cluster is dynamically increased with transient nodes allocated through the Grid Engine workload manager. The objective was to support ‘bursty’ Kubernetes workloads which may require HPC computing resources while also maintaining a uniform platform with a homogeneous job management and monitoring infrastructure.  López-Huguet et al. \cite{inbook} propose a cloud job scheduler which can manage and submit jobs in a HPC infrastructure. This implementation is similar to the Bridge Operator but is based on slightly different concepts. In López-Huguet et al.  all functionality  is contained inside one  image which runs in a Kubernetes pod. In Cerin et al. \cite{9235080} the approach is to containerise the HPC scheduler itself making it cohabit with all other jobs on the underlying cloud system. This approach represents a direct integration of Cloud and HPC environments and is in contrast to our black box approach to deploying jobs on an external system. Georgiou et al. \cite{Georgiou} focused on hybrid data analytics workflows in the context of precision agriculture and livestock farming applications and implemented a prototype architecture which integrates with Kubernetes with a HPC partition of baremetal nodes managed by Slurm or Torque.
Recently the DOE's Oak Ridge Leadership Computing Facilities (OLCF) extended the  Pegasus workflow management system with Workflow Submit Node as a service (WSaaS). Built upon the Kubernetes/Openshift cluster (Slate) that exists within OLCF’s DMZ this service  allows users  to spin up a Pegasus submit node, and submit pipelines to OLCF resources such as the Summit LSF cluster or Rhea Slurm Cluster \cite{osti_1665960}.

To the best of the authors’ knowledge, the Bridge Operator approach presented in this work is novel in two key respects where
\begin{itemize}
    \item  the operator mirrors the external resource in pods, which act as proxies to control the external system. The Bridge operator controls and monitors these pods and in effect controls and monitors the external resource by proxy.
\item the operator represents a pattern with a generic approach to control multiple different external systems, with  the same programming  model.
\end{itemize}

\begin{figure}
\includegraphics[width=8cm]{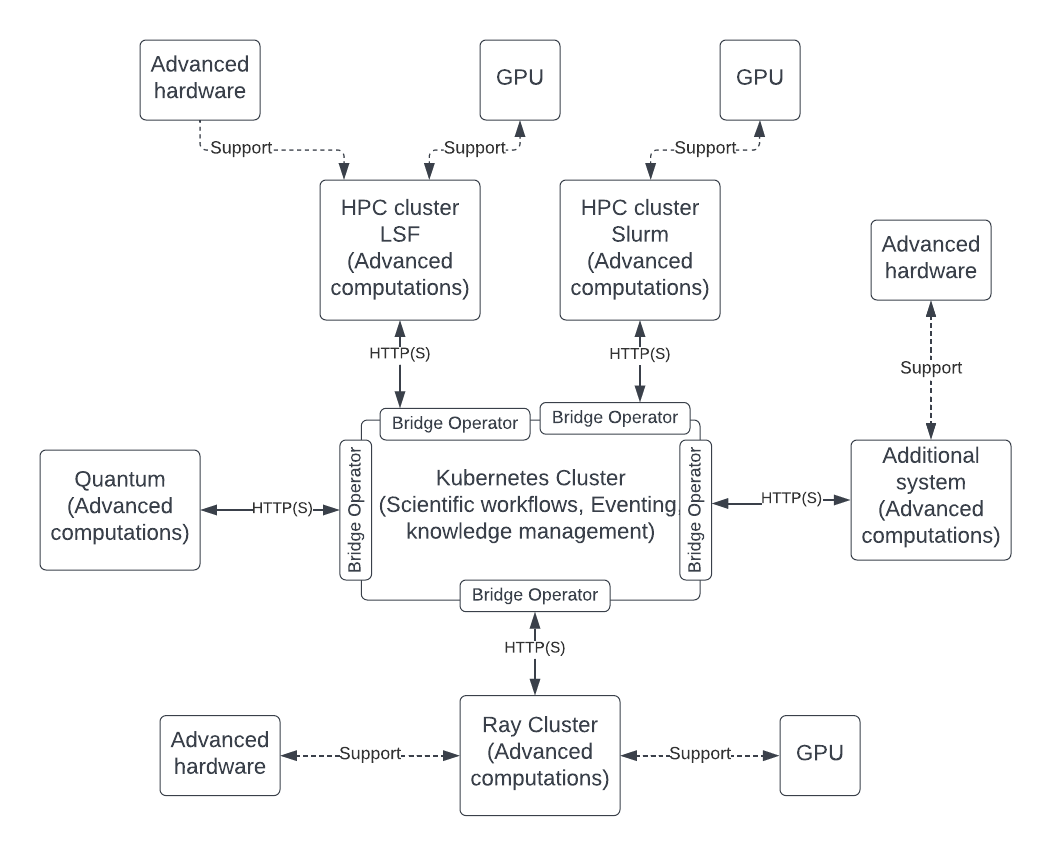}\label{fig:highlevelarch}
\caption{The Bridge Operator running workflows on a Kubernetes Cluster while at the same time deploying workflow steps (complex computations) to an external system (e.g. HPC cluster, Quantum service or Ray cluster) managed by an external resource manager.}
\centering
\end{figure}

\section{Objectives} \label{sect:obj}
The goal in creating the Bridge Operator is to support running complex workflows on the cloud while at the same time seamlessly deploying workflow steps (complex computations) to an external system (e.g. HPC cluster, Quantum service or Ray cluster) managed by an external resource manager as shown in Fig \ref{fig:highlevelarch}.
In our implementation we use a single Kubernetes Operator which is capable of deploying pods which can simultaneously submit and monitor jobs on different external systems as required in the workflow. The Bridge operator watches pods adhering to the following general requirements:
\begin{itemize}
    \item The ability to submit user provided credentials to access the external resource via a HTTP/HTTPS API.
\item The ability to submit, delete and query the status of a job running on an external resource via the API.
\item The ability to submit user provided job property settings, such as accounts to charge node usage to, queues to run on as well as hardware resource requests, to the external workload manager 
\item The ability to upload execution scripts, and any additional files needed to the external resource as allowed by the API..
\item The ability to download any output files from a given location on the external resource. 
\item Using S3 (compatible) storage for all uploads/downloads. 
\item Credentials to access the external resources as well as object storage are accessible as Kubernetes secrets mounted in a volume by the pod.
\end{itemize}

\section{Architecture and Tools}\label{sect:archntools}
In this section we describe the Operator Architecture which has been developed to run on a Kubernetes Cluster and deploy a long running application to a HTTP/HTTPS API on an external resource. We outline the operator’s controller, CR, controller logic and credential management. Specific implementations for a LSF, Slurm, Quantum and Ray Pod are also presented. 

\begin{figure}
\includegraphics[width=8cm]{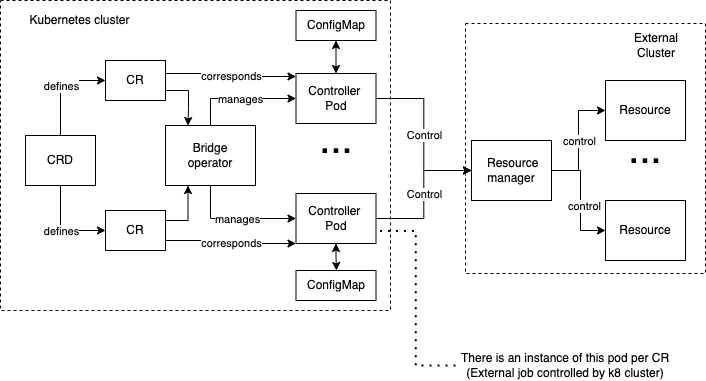}\label{fig:lowlevelarch}
\caption{A detailed view of our implementation and the control, management and definition relationships within the Bridge Operator.}
\centering
\end{figure}

\subsection{Bridge Operator Architecture}
In Fig. \ref{fig:highlevelarch} we presented a view of the Bridge Operator which allows jobs on a Kubernetes cluster to be submitted to an external system.  Fig \ref{fig:lowlevelarch} presents a detailed view of our implementation and the control, management and definition relationships within the Bridge Operator. For each resource type supported by the operator (LSF cluster, Slurm cluster, Quantum, Ray) we have created a controller, specialized for this type of resource, responsible for all interactions with this resource type. The Bridge operator itself is responsible for the management of controller lifecycles. While the operator is generic, implementation of a controller pod is specific for a given external resource manager. As a result, in order to support a new resource type, the only thing that is required is the implementation of the corresponding controller (based on very simple rules imposed by the operator).
An operator is controlled by a custom resource (CR), that defines parameters for the execution, including:
\begin{itemize}
    \item The resource manager URL and security credentials for accessing it
\item The controller pod used for controlling the resource (docker image)
\item The job execution script and its location - the operator supports `remote` (located at the external resource), `s3` (located in an S3 bucket) and `inline` (defined by the CR itself) scripts.
\item Any additional job data and its location. For example input files may be uploaded from S3 to the external resource.
\item Any job parameter settings required by the job execution script. The operator supports both S3-based and inline specification of this data
\item Any job properties specifications, for example, specifying resources (number of nodes), execution time, error and output files, etc.
\item S3 information (s3 endpoint and credentials)
\item S3 upload specifications (s3 bucket and files to be uploaded)
\end{itemize}
Additionally the Bridge operator uses a configuration map for passing data between the operator and controller pod and as a persistent current record of the job’s status should the pod fail. Note that in this implementation we focus on S3 for simplicity but the approach can be generalized to other storage options.
Once the CR is created, an operator creates the controller pod (one per remote job) and populates the configuration map with the parameters required for the pod’s execution. Once the pod is created, an operator watches for changes in the pod and associated configuration map. Once execution of the remote job is completed a CR status reflects the status of remote job as any of DONE/KILLED/FAILED/UNKNOWN as well as additional information about the job start and end time. A user can also update the CR with a kill signal which will stop the remote job and update the status to KILLED. When a CR is deleted, it will clean up all resources associated with it (pod and configuration map)
The workhorse of this implementation is the controller pod. Once the pod is created, it reads the necessary execution data from the associated configuration map, logs into a remote resource manager, fetches a job script (if required) along with job parameters and any other additional data and submits the requested job via the workload manager of the resource. After the job starts, the pod goes into a monitoring loop, that runs periodically (controlled by the CR poll parameter) checking the job status and updating the associated configuration map. Once execution of the remote job completes/fails or is killed, the pod downloads outputs to the Kubernetes cluster and uploads them to S3, if required, and exits. If the pod fails for any reason, an operator will restart it. Because the remote job ID is kept in the config map, in this case the pod will know that the remote job is already running and will not try to restart it.
Credentials necessary for deploying jobs and accessing object storage (if required) are managed using Kubernetes secrets. A specific Bridge Job can be deployed using a CR yaml file, see Fig. \ref{slurmyaml} for example yaml.

\begin{algorithm}
\floatname{algorithm}{}
\caption{Sample BridgeJob yaml for submitting to a Slurm cluster}\label{slurmyaml}
\begin{algorithmic}[1]
\State kind: BridgeJob
\State apiVersion: bridgeoperator.ibm.com/v1alpha1
\State metadata:
\State \qquad name: slurmjob-test
\State spec:
\State \qquad  resourceURL: http://my-slurm-cluster@hpc.com
\State \qquad image: slurmpod:0.1
\State  \qquad resourcesecret: mysecret
\State \qquad imagepullpolicy: Always
\State \qquad updateinterval: 20
\State \qquad jobdata:
\State  \qquad \qquad  jobscript: "mys3bucket:slurmbatch.sh"
\State  \qquad \qquad  scriptlocation: "s3"
 \State \qquad \qquad jobproperties: $|$ \\
 \State \qquad \qquad \qquad    \{
 \State  \qquad \qquad \qquad   "NodesNumber":"1", "Queue": "V100", "Tasks": "2", "slurmJobName": "test",
 \State   \qquad \qquad \qquad  "currentWorkingDir": "path-to-test/test-script/",
\State   \qquad \qquad \qquad   "envLibPath": "/usr/mpi/gcc/openmpi-4.0.3rc4/lib",
 \State  \qquad \qquad \qquad   "ErrorFileName": "slurmjob.err",
 \State  \qquad \qquad \qquad   "OutputFileName": "slurmjob.out"
 \State  \qquad \qquad \qquad   \}
\State \qquad s3storage:
 \State \qquad  s3secret: mysecret-s3
 \State \qquad  endpoint: s3endpoint.cloud
 \State \qquad  secure: false 
 \end{algorithmic}
\end{algorithm}

\subsection{Bridge Pods}
The specific implementation for a given external resource is carried out by a Pod which is specific for each kind of workload managers, e.g. the LSF or Slurm. To demonstrate the generic approach of the Bridge operator we have created pods which can submit jobs to an LSF, Slurm, Ray cluster or to a Quantum service. 
Any actions within the pod itself are restricted to what can be done via specific external resource API. For example the LSF Application Center API currently allows requests to submit, delete, update jobs, upload and download files to the cluster as well as query queues and job statuses. The current version of the Slurm API tested in this work (version 21.08) is slightly more restrictive in that upload and download to the cluster is not supported.
There is a common sequence of events within a pod which we summarize in Figs. \ref{podmain},\ref{podmonitor} for a pod implemented in the Go programming language \cite{golang}.
Fig \ref{podmain} outlines the main method in the controller pod.
Fig \ref{podmonitor} outlines the monitoring method which runs continuously in the controller pod.

\begin{algorithm}
\floatname{algorithm}{}
\caption{Pseudocode for the \emph{main} function submitting jobs to a Slurm HPC cluster}\label{podmain}
\begin{algorithmic}[1]
\State func main()\{
\State \qquad	NAMESPACE $=$ os.Getenv$($``NAMESPACE"$)$ $//$Get namespace, job name from environment
\State	\qquad JOB\_NAME $=$ os.Getenv$($``JOBNAME"$)$
\State	\qquad config, err $:=$ rest.InClusterConfig$()$  $//$  Create kubernets client
\State \qquad $\cdots$ 
\State \qquad clientset, err $:=$ kubernetes.NewForConfig$($config$)$
\State \qquad	cm := getConfigMap$($clientset$)$ $//$Get config map and its parameters
\State \qquad HPCURL $=$ cm.Data[``resourceURL"]
\State \qquad $\cdots$
\State	\qquad client $=$ http.Client\{Timeout: time.Duration$($5$)$ * time.Second\} $//$ Create HTTP client
\State \qquad 	slurmUsername, slurmToken  $:=$ getToken$()$. $//$ Get access token for HPC cluster
\State \qquad $\cdots$ 
\State \qquad	id $:=$ cm.Data[``id"]  $//$ Get ID from config map
\State	\qquad info $:=$ make$($map[string]string$)$ $//$ create info for keeping track of execution parameters
\State	\qquad $\cdots$
\State	\qquad if len$($id$)$ $==$ 0 \{ 
\State \qquad \qquad		klog.Infof$($``Slurm Job with name '\%s' does not exist. Submitting new job.", JOB\_NAME$)$
\State \qquad \qquad		intId $:=$ submit$($slurmUsername, slurmToken, cm.Data$)$ $//$ Submit a job \& update execution state in config map
\State \qquad \qquad id = fmt.Sprint(intId)
\State \qquad \qquad		if id == 0 \{
\State \qquad \qquad \qquad 			info[``jobStatus"] $=$ FAILED 	$//$  Failed to submit a job
\State \qquad\qquad \qquad 			info[``message"] $=$ ``Failed to submit a job to HPC resource"
\State \qquad\qquad 		\} else \{
\State \qquad \qquad \qquad 			info[``id"] $=$ id
\State \qquad \qquad \qquad 			info[``jobStatus"] $=$ SUBMITTED
\State \qquad \qquad 		\}
\State \qquad \qquad 		updateConfigMap$($clientset, cm, info, id$)$
\State \qquad \qquad 		if len$($id$)$ $!=$ 0 \{ $//$ Start monitoring or exit
\State \qquad \qquad \qquad			monitor$($clientset, slurmUsername, slurmToken, info$)$
\State \qquad \qquad		\} else \{
\State \qquad \qquad \qquad			klog.Exit$($``Failed to start HPC job"$)$
\State \qquad\qquad		\}
\State \qquad \} else \{
\State \qquad\qquad		$//$ Job is already running
\State\qquad \qquad		klog.Info$($``Slurm Job '\%s' has ID in ConfigMap. Handling state.", JOB\_NAME$)$
\State \qquad \qquad		info[``id"] $=$ id
\State \qquad \qquad  monitor$($clientset, slurmUsername, slurmToken, info$)$
\State \qquad \qquad 		\}
\State \}
\end{algorithmic}
\end{algorithm}

\section{Workflow Integration}\label{sect:workflow}
Although Bridge operator is useful by itself, the majority of scientific workflows are implemented using a specialized workflow package. The one that we used, Kubeflow pipelines is a platform for deploying workflows based on Docker images and provides a useful SDK for defining and controlling components and pipelines. To demonstrate a workflow integration with presented  Bridge operator implementation we have created a single pipeline which consists of three steps as shown in Fig. \ref{bridgepipeline}: create a configuration map (createop), run a Bridge pod (invokeop) and delete the configuration map once the job is complete (cleanop). Fig \ref{bridgepipeline} shows the python code to create this pipeline. The input parameters to the bridge pipeline correspond to the BridgeJob schema defined for the Operator. In particular the parameter “docker, str” should reference a docker image which is a Bridge Pod to submit jobs to a specific external resource.
Depending on the parameters, this workflow implementation can be used with any of the Bridge operator pods.

\begin{algorithm}
\floatname{algorithm}{}
\caption{Pseudocode for the \emph{monitor} function}\label{podmonitor}
\begin{algorithmic}[1]
\State func monitor(clientset *kubernetes.Clientset, slurmUsername string, slurmToken string, info map[string]string)\{
\State  id := info[``id"]
\State for \{
\State \qquad		time.Sleep$($time.Duration$($POLL$)$ * time.Second$)$
\State \qquad $//$ Get current config map
\State \qquad		cm := getConfigMap$($clientset$)$

\State \qquad		$//$ Get current execution status and update config map
\State \qquad		var state = ``"
\State \qquad		job := getJobInfo$($slurmUsername, slurmToken, id$)$
\State \qquad if job $!=$ nil \{
\State \qquad \qquad $\cdots$
\State \qquad \qquad updateConfigMap$($clientset, cm, info, token, id$)$
\State \qquad  \}
\State \qquad $//$ Check for kill flag
\State \qquad		if cm.Data[``kill"] $==$ ``true" \{
\State \qquad \qquad			killJob$($slurmUsername, slurmToken, id, info[``jobStatus"], info$)$
\State \qquad		\}
\State \qquad $//$ Terminate if we are done
\State \qquad if state $==$ COMPLETED \{
\State \qquad \qquad			os.Exit$($0$)$
\State \qquad		\}
\State \qquad if state $==$ CANCELLED $||$  state $==$ FAILED \{
\State \qquad \qquad			os.Exit$($1$)$
\State \qquad		\}
\State \} 
\State \}
\end{algorithmic}
\end{algorithm}

\begin{algorithm}
\floatname{algorithm}{}
\caption{Pseudocode for the KFP Bridge pipeline}\label{bridgepipeline}
\begin{algorithmic}[1]
\State def bridgepipeline$($jobname: str,                              \# job name
 \State \qquad   \qquad             namespace: str,                            \# execution namespace
\State \qquad	\qquad         resourceURL: str,             \# resource address - url
\State \qquad    \qquad             resourcesecret: str,          \# resource credentials
\State \qquad    \qquad             script: str,                  \# script name or content
\State \qquad    \qquad             scriptlocation: str,  
\State \qquad    \qquad             docker: str,                            \# pod docker name
 \State \qquad   \qquad             scriptmd: str $=$ "",
 \State \qquad    \qquad            scriptextraloc: str$=$ "",
\State \qquad     \qquad            additionaldata: str $=$ "",          \# extra files required
\State \qquad     \qquad            jobproperties: str $=$ "",           \# dict of job properties
\State \qquad     \qquad            jobparams: str $=$ "",           \# dict of job parameters
 \State \qquad    \qquad            s3secret: str $=$ "",                \# secret with S3 credentials
 \State \qquad    \qquad            s3endpoint: str $=$ "",              \# S3 URL
 \State \qquad    \qquad            s3secure: str $=$ "",           \# is S3 secure?
 \State \qquad     \qquad           s3uploadfiles: str $=$ "",           \# files to upload to S3
 \State \qquad     \qquad           s3uploadbucket: str $=$ "",           \# bucket in S3
 \State \qquad     \qquad           updateinterval: str $=$ "20",             \#  poll interval
\State \qquad	\qquad	 imagepullpolicy: str $=$ ``IfNotPresent"
\State \qquad     \qquad           $ )$:
 \State \qquad   createop $=$ setup\_op$($jobname, namespace, resourceURL, resourcesecret, script, 
 \State \qquad \quad scriptlocation,scriptmd, additionaldata, scriptextraloc, jobproperties, jobparams, \
 \State \qquad \quad                   s3secret, s3endpoint, s3secure, s3uploadfiles, s3uploadbucket,updateinterval$)$
 \State \qquad   createop.execution\_options.caching\_strategy.max\_cache\_staleness $=$ ``P0D"
 \State \qquad    invokeop $=$ comp.load\_component\_from\_text$($"""
 \State \qquad   \qquad     name: bridge-pod
 \State \qquad   \qquad     description: bridge execution pod
  \State \qquad  \qquad     implementation:
  \State \qquad    \qquad       container:
   \State \qquad   \qquad           image: docker
   \State \qquad   \qquad           command:
   \State \qquad    \qquad          - sh  """$)$$($$)$ \
  \State \qquad    \qquad   .add\_volume(k8s\_client.V1Volume(name$=$'credentials',
   \State \qquad \qquad secret=k8s\_client.V1SecretVolumeSource(secret\_name$=$resourcesecret))) \
   \State \qquad  \qquad    .add\_volume\_mount(k8s\_client.V1VolumeMount(mount\_path$=$'/credentials',name$=$'credentials')) \
  \State \qquad  \qquad    .add\_env\_variable(k8s\_client.V1EnvVar(name$=$'NAMESPACE', value$=$namespace)) \
  \State \qquad   \qquad   .add\_env\_variable(k8s\_client.V1EnvVar(name$=$'JOBNAME', value$=$jobname)) \
  \State \qquad  \qquad    .after(createop)
  \State \qquad  invokeop.container.set\_image\_pull\_policy(imagepullpolicy)
  \State \qquad  invokeop.container.image $=$ docker
 \State \qquad   invokeop.execution\_options.caching\_strategy.max\_cache\_staleness $=$ "P0D"
 \State \qquad   if s3secret $!=$ ``":
  \State \qquad   \qquad   invokeop.add\_volume(k8s\_client.V1Volume(name$=$'s3credentials',
 \State \qquad   \qquad       secret=k8s\_client.V1SecretVolumeSource(secret\_name$=$s3secret))) \
 \State \qquad    \qquad       .add\_volume\_mount(k8s\_client.V1VolumeMount(mount\_path$=$'/s3credentials', \State \qquad    \qquad  \qquad name='s3credentials'))
 \State \qquad   invokeop.add\_volume(k8s\_client.V1Volume$($name$=$'downloads'$)$$)$
 \State \qquad     invokeop.add\_volume\_mount(k8s\_client.V1VolumeMount(mount\_path='/downloads', 
 \State \qquad  \qquad  \qquad name$=$'downloads'$)$$)$
\State \qquad    cleanop $=$ cleanup\_op$($jobname, namespace$)$.after$($invokeop$)$
\State \qquad    cleanop.execution\_options.caching\_strategy.max\_cache\_staleness $=$ "P0D"
\end{algorithmic}
\end{algorithm}

\section{Conclusions}\label{sect:conc}
There is a growing trend for complex scientific workflows which require heterogeneous computing resources, combining cloud with HPC and even advanced accelerators such as AI and quantum.  To address these needs we have created a Kubernetes ‘Bridge’ Operator which allows jobs to be deployed and monitored on remote external resources from a cloud Kubernetes environment. The black box approach taken in this implementation has resulted in a generic pattern which works for different external resources (Slurm, LSF, Quantum, Ray, etc) without any change to the operator required. The specific requests made to the external resource are restricted to the worker pod controlled and monitored by the operator. In this implementation the external resource is mirrored in the pod, which acts as a proxy for that system. As the Bridge Operator controls the pod, it in turn is effectively controlling the external system by proxy.
We demonstrate the workflow integration of this approach in Kubeflow pipelines as a simple three step workflow, which can be used as a sub workflow for more complex implementations.
The operator presented here allows a user to manage jobs on remote resources, but does not take into the account the current load on these systems. Future work will focus on creating  companion operator using the same approach to monitor current load on these remote resources and make intelligent decisions on which remote resource (or particular portion of this resource, for example specific queue) to use for execution.


\begin{backmatter}

\section*{Acknowledgements}
The authors would like to thank Dean Wampler for a detailed reading of this manuscript and feedback.

\section*{Ethical Approval and Consent to participate}
Not applicable
\section*{Consent for publication}
Not applicable
\section*{Availability of supporting data}
Not applicable
\section*{Competing interests}
The authors declare that they have no competing interests.
\section*{Funding}
Not applicable
\section*{Authors' contributions}
All authors contributed to writing and testing the code, as well as writing the text of this paper.


\bibliographystyle{bmc-mathphys} 
\bibliography{main}      





\end{backmatter}
\end{document}